\documentclass[aps, prl, reprint, twocolumn, superscriptaddress, amsmath, amssymb, floatfix,nobibnotes]{revtex4-2}
\usepackage{graphicx}
\usepackage{dcolumn}
\usepackage{bm}
\usepackage{enumerate}
\usepackage{fancyhdr}
\usepackage{color}
\usepackage[normalem]{ulem}
\usepackage[squaren]{SIunits}
\usepackage[mathlines]{lineno}
\usepackage[colorlinks=true,linkcolor=blue]{hyperref}
\hypersetup{
    citecolor=blue,linkcolor=blue
}

\begin{document}

\title{TeV Electron Beams from Plasma Acceleration via Regenerative Cascading}

\author{Chaojie Zhang}
\email[Contact author: ]{chaojiez@ucla.edu}
\affiliation{Department of Electrical and Computer Engineering, University of California, Los Angeles, Los Angeles, CA, USA}
\author{Chan Joshi}
\affiliation{Department of Electrical and Computer Engineering, University of California, Los Angeles, Los Angeles, CA, USA}

\date{\today}

\begin{abstract}
Plasma accelerators sustain gradients orders of magnitude higher than conventional radiofrequency machines, but most proposed paths to TeV energies still require tens of stages, each demanding sub-micrometer alignment, femtosecond synchronization, and precise matching of the accelerating trailing bunch. Here we introduce plasma wakefield acceleration via regenerative cascading, in which each stage self-injects a fresh trailing electron bunch and the accelerated trailing bunch becomes the driver for the next stage. This approach has several advantages: energy multiplication instead of addition; automatic alignment, synchronization, and matching of the trailing bunch to the wake; and trailing bunch brightness reset in each stage. Particle-in-cell simulations show the generation of a 1.1 TeV electron beam with $\sim$0.3\% rms energy spread and 0.12 nC charge from a two-stage, sub-kilometer plasma accelerator driven by a 45 GeV, 100 nC beam. The low energy spread is achieved via dynamic beam loading in the evolving wake of the post-depletion driver that acts as a built-in energy dechirper.
\end{abstract}
\maketitle

Beam-driven plasma wakefield acceleration (PWFA) has made rapid progress in the past several decades~\cite{chen_acceleration_1985,joshi_plasma_2006,joshi_perspectives_2020}. PWFA experiments have demonstrated energy gains exceeding 40 GeV in a meter-scale plasma~\cite{blumenfeld_energy_2007}, high energy-transfer efficiency with beam loading~\cite{litos_high-efficiency_2014}, and preservation of energy spread at the sub-percent level~\cite{lindstrom_energy-spread_2021} and emittance at the mm-mrad level~\cite{lindstrom_emittance_2024}. Most recently, a plasma wakefield accelerator has been shown to simultaneously boost beam energy and brightness through downramp injection~\cite{zhang_plasma-wakefield_2025}, which serves as the foundation and motivation for the present work.

Despite this progress, the most challenging application for plasma accelerators, a linear collider at the energy-frontier of particle physics that requires ultra-bright TeV energy electron beams~\cite{schroeder_physics_2010,adli_plasma_2019,gessner_design_2025}, remains out of reach. Most proposed PWFA-based colliders share a common architecture for the electron arm. A trailing electron bunch injected behind the drive bunch at the start of the first stage gains energy while depleting the drive bunch. This process is thereafter repeated through many stages, increasing the energy of the trailing bunch linearly while preserving its quality. However, preserving the beam quality through tens of such stages demands sub-micrometer alignment and femtosecond synchronization between the drive and the trailing bunch, and matching of the trailing bunch at the interface of every stage~\cite{adli_plasma_2019}. Misalignment can trigger the hosing instability that degrades beam quality or even causes charge loss~\cite{whittum_electron-hose_1991,huang_hosing_2007}; synchronization errors lead to a jitter in the positioning of the trailing bunch at an unintended wake phase and thereby cause an energy jitter~\cite{lindstrom_staging_2021}; and an unmatched beam undergoes envelope oscillations that degrade emittance~\cite{mehrling_transverse_2012,lindstrom_staging_2016}. These are long-standing challenges~\cite{adli_plasma_2019} that require fresh solutions.

A separate challenge is the reduction of average accelerating gradient by interstage beam optics~\cite{lindstrom_staging_2016}. While plasma wakes often sustain 10-100 GV/m gradients, the beam optics required between stages (to remove the spent driver, inject a fresh one, and match the trailing bunch) increase in length as the beam energy increases, thereby reducing average accelerating gradients to less than 1 GV/m in current multi-stage designs~\cite{foster_hybrid_2023}. Although solutions to each of these issues have been proposed~\cite{knetsch_high_2023}, no single design addresses all of them, to our knowledge.

In this Letter, we introduce a new PWFA scheme we call Plasma Acceleration via Regenerative Cascading (PARC). Rather than accelerating the same trailing bunch through many stages and simultaneously preserving its quality, PARC produces a trailing bunch by self-trapping plasma electrons in each stage, and uses the accelerated trailing bunch as the driver for the subsequent stage. This provides several key advantages. First, each trailing bunch is self-injected and is therefore automatically aligned, synchronized, and approximately matched to the wake, eliminating the stringent staging tolerances that have limited conventional approaches~\cite{adli_plasma_2019}. Second, trailing bunch energy accumulates multiplicatively rather than additively over stages, dramatically reducing the number of stages (from tens to just two) required to reach TeV energies. Third, the brightness of the trailing bunch is determined by the injection process in each stage, and can thus increase rather than degrade over stages. Moreover, interstage optics is greatly reduced, restoring a high average gradient (e.g., $>1$ GV/m). Start-to-end particle-in-cell (PIC) simulations demonstrate that a 45 GeV, 100 nC driver produces a 1.1 TeV electron bunch with 0.12 nC charge, $\sim$0.3\% projected rms energy spread (0.02\% slice), $\sim$2.4 mm$\cdot$mrad normalized emittance, and multi-kA peak current in two plasma stages with a total plasma length of 825 meters.

\begin{figure}[tb]
\centering
\includegraphics[width=\columnwidth]{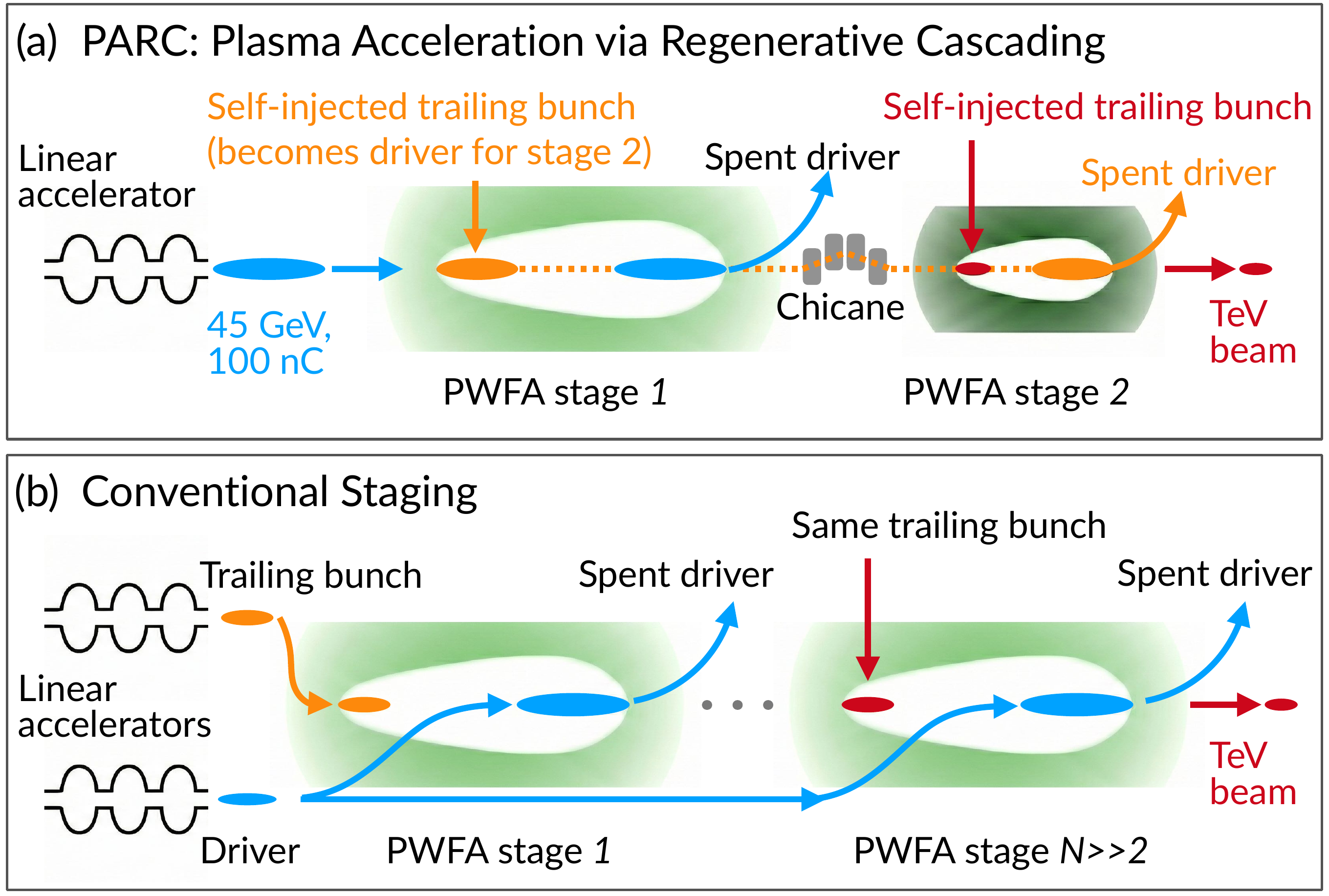}
\caption{(a) The PARC (Plasma Acceleration via Regenerative Cascading) concept. A high-charge beam (e.g., 45 GeV, 100 nC) drives the first PWFA stage, where a trailing bunch is self-injected from the plasma and gets accelerated. After the first stage, the trailing bunch is compressed by a magnetic chicane and then serves as the driver for the second stage, where a fresh trailing bunch is again self-injected and accelerated to TeV energy. (b) Conventional staging concept, in which the same trailing bunch traverses many ($N\gg2$) stages, each requiring a fresh driver and precise alignment, synchronization, and matching at every interface.}
\label{fig:concept}
\end{figure}

Figure \ref{fig:concept}(a) illustrates the PARC concept. The first stage is driven by a high-charge, relatively low-energy bunch from a conventional accelerator. The driver's current profile is shaped to achieve high-transformer-ratio acceleration~\cite{su_optimization_2023,loisch_observation_2018,roussel_single_2020}. A trailing bunch is created inside the wake via controlled injection (e.g., density downramp injection~\cite{bulanov_particle_1998,suk_plasma_2001,buck_shock-front_2013} or ionization injection~\cite{pak_injection_2010}). After injection, the trailing bunch gains energy as the driver decelerates over the rest of the plasma. Upon exit, the spent driver is removed by a dipole magnet using the large energy difference between the depleted driver and the accelerated trailing bunch. The trailing bunch underloads the wake~\cite{tzoufras_beam_2008} so that it develops a large, nearly linear energy chirp, and can therefore be compressed by a modest-size chicane with minimal emittance growth induced by coherent synchrotron radiation (CSR)~\cite{emma_terawatt_2021}. After compression, the trailing bunch is sent into the second stage and now acts as the driver to repeat the process whereby a fresh trailing bunch is once again self-injected and accelerated.

Each stage can be characterized by the energy transformer ratio $R_i\equiv\Delta E_T/\Delta E_D$, where $\Delta E_T$ is the trailing bunch energy gain and $\Delta E_D$ is the driver energy loss. When the driver approaches pump depletion, $\Delta E_D$ approaches its initial energy, so $R_i\approx E_T/E_D$. The trailing bunch energy after $N$ stages is therefore
\begin{equation}
E_N = E_0 \prod_{i=1}^{N} R_i,
\label{eq:energy}
\end{equation}
where $E_0$ is the initial driver energy. With a driver-to-trailing energy conversion efficiency $\eta_i$ in each stage, the maximum achievable trailing bunch charge after $N$ stages is
\begin{equation}
Q_N = Q_0 \prod_{i=1}^{N} \eta_i R_i^{-1},
\label{eq:charge}
\end{equation}
where $Q_0$ is the initial driver charge.

Three consequences follow. First, energy accumulates multiplicatively, so TeV energies can be reached in just a few stages using a tens of GeV driver for $R_i\gg1$. Second, a high initial driver charge is necessary to deliver useful trailing charge at the last stage. Third, the overall energy efficiency scales as $\eta^N$ rather than $\eta$ for conventional staging at the same per-stage efficiency (e.g. $\eta\sim0.4$~\cite{litos_high-efficiency_2014,joshi_plasma_2018}), favoring small $N$.

PARC therefore differs from the conventional staging concept shown in Fig. \ref{fig:concept}(b) by transferring the energy of a single high-charge driver through a regenerative chain of PWFA stages, rather than distributing it among multiple identical fresh drivers for independent stages. This eliminates the stringent staging tolerances that have limited conventional approaches and enables more rapid trailing bunch energy gain, at the cost of the $\eta^N$ efficiency.

\begin{figure*}[tb]
\centering
\includegraphics[width=0.95\textwidth]{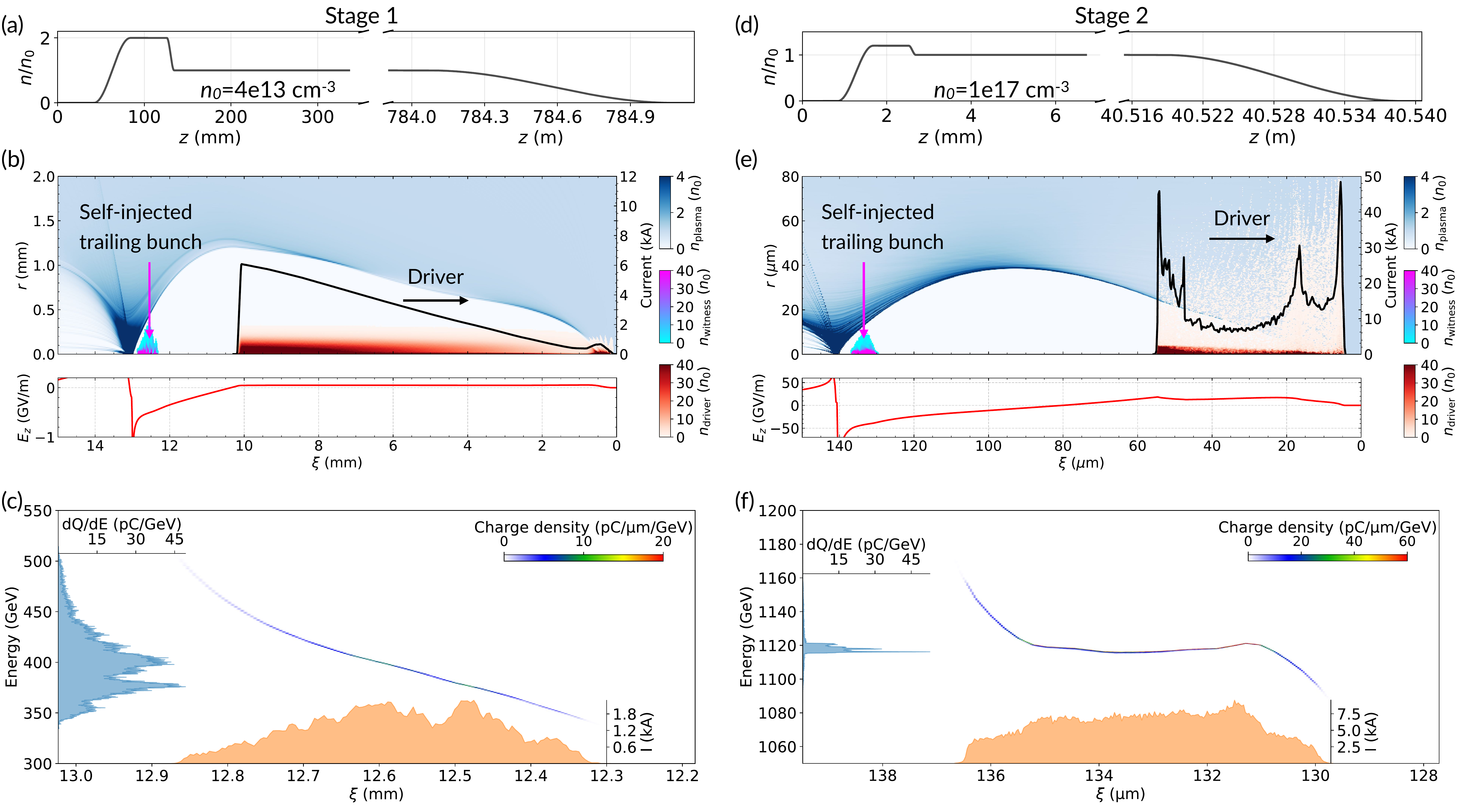}
\caption{Two-stage PARC simulation results. Stage 1 (a-c): (a) Plasma density profile: a 2:1 plasma density ratio downramp over 8.4 mm near the entrance of a 785 m uniform region having $n_0=4\times10^{13}$ $\rm cm^{-3}$, and a $\sim$10 cm exit downramp. (b) Snapshot of the wake $\sim$10 cm after injection, showing plasma electron density (blue), the shaped driver (orange, with current profile in black), the downramp-injected trailing bunch (magenta), and the on-axis electric field $E_z$ (red, lower panel). The 45 GeV, 100 nC driver experiences a uniform decelerating field of $\sim$50 MV/m whereas the downramp-injected trailing bunch accelerates at $\sim$0.5 GV/m, corresponding to a voltage transformer ratio of 10. (c) Longitudinal phase space of the trailing bunch at the stage exit. The bunch contains 2.3 nC total charge and reaches 400 GeV mean energy with a large quasi-linear energy chirp suitable for compression. Energy spectrum (left) and current profile (bottom) are projections from the phase space. Stage 2 (d-f): (d) Plasma density profile: a 1.2:1 density ratio downramp over 168 $\mu$m near the entrance and a 40 m uniform region at $n_0=10^{17}$ $\rm cm^{-3}$. (e) Wake driven by the compressed trailing bunch from stage 1, with a second downramp-injected trailing bunch accelerating at a loaded gradient of $\sim$40 GV/m. (f) Longitudinal phase space of the final accelerated trailing bunch: 1.1 TeV peak energy, 0.3\% projected rms energy spread, and 0.12 nC charge.}
\label{fig:stages}
\end{figure*}

We demonstrate the PARC concept using start-to-end PIC simulations combining two codes: OSIRIS~\cite{goos_osiris_2002} and QPAD~\cite{li_quasi-static_2021}. The former is a fully relativistic, multi-dimensional code that resolves the injection dynamics, and the latter uses the quasi-static approximation for efficient modeling of long-distance acceleration. Downramp injection and pre-acceleration to $>$100 MeV are modeled in OSIRIS; the resulting 6D phase space of the driver and trailing bunch are then imported into QPAD for the subsequent acceleration. Simulation details are described in Appendix A.

\textit{Stage 1}. The first stage uses a 45 GeV, 100 nC bunch with an approximately triangularly ramped current profile rising from head to tail over $\sim$33 ps and peaking at 6.2 kA [see the black line in Fig.~\ref{fig:stages}(b)] to drive the PWFA in a plasma with the density profile shown in Fig. \ref{fig:stages}(a). The driver's current profile is optimized using the method of Ref.~\cite{su_optimization_2023} so that it excites a nonlinear wake~\cite{Lu:2006ci,lu_generating_2007} in which the entire bunch decelerates at nearly the same rate, enabling high-transformer-ratio acceleration [Fig. \ref{fig:stages}(b)]. As the driver traverses the sudden density downramp, the abrupt elongation of the wake traps plasma electrons, forming a fresh trailing bunch that is then accelerated over the rest of the plasma. At the stage exit [Fig. \ref{fig:stages}(c)], the trailing bunch reaches $\sim$400 GeV mean energy with 2.3 nC total charge, a peak current of 2.4 kA over $\sim$120 $\mu$m rms length ($\sim$500 $\mu$m full length), a normalized emittance of $\sim$160 mm$\cdot$mrad, and a large, nearly linear energy chirp suitable for compression. On its own, this beam already exceeds the highest electron energy previously produced in the laboratory and could enable strong-field quantum electrodynamics (SFQED) experiments in the $\chi\gg1$ regime~\cite{yakimenko_prospect_2019,mirzaie_all-optical_2024}.

\textit{Interstage compression}. The large, quasi-linear energy chirp accumulated in the under-loaded wake of stage 1 allows the accelerated 400 GeV trailing bunch to be compressed in a modest-strength magnetic chicane despite its high energy. Calculations show that a chicane with $R_{56}=-1.54$ mm compresses the bunch length to $\sim$50 $\mu$m (full length), producing an average current of 20 kA suitable for driving a nonlinear wake at higher plasma density in stage 2. The high beam energy and small $|R_{56}|$ strongly suppress the CSR effects~\cite{swanson_longitudinal_2026}; an analytic estimate predicts MeV-level energy spread increase and negligible emittance growth (Appendix B).

\textit{Stage 2}. The compressed trailing bunch drives the second stage in a plasma of $n_0=10^{17}$ $\rm cm^{-3}$, 2500 times denser than stage 1, enabled by the 200 times shorter driver. A smaller 1.2:1 density ratio downramp [Fig. \ref{fig:stages}(d)] injects a new, short trailing bunch close to the back of the wake to maximize energy gain. Despite a spiky driver current profile resulting from the energy chirp and compression nonlinearities [Fig. \ref{fig:stages}(e)], the decelerating field remains roughly flat at $\sim$20 GV/m. The new trailing bunch is accelerated at $\sim$40 GV/m, yielding a voltage transformer ratio of 2. After 40 m, the trailing bunch reaches 1.1 TeV peak energy with 0.3\% projected (0.02\% slice) rms energy spread, 0.12 nC charge, and $\sim$8 kA peak current [Fig. \ref{fig:stages}(f)].

\begin{figure}[tb]
\centering
\includegraphics[width=\columnwidth]{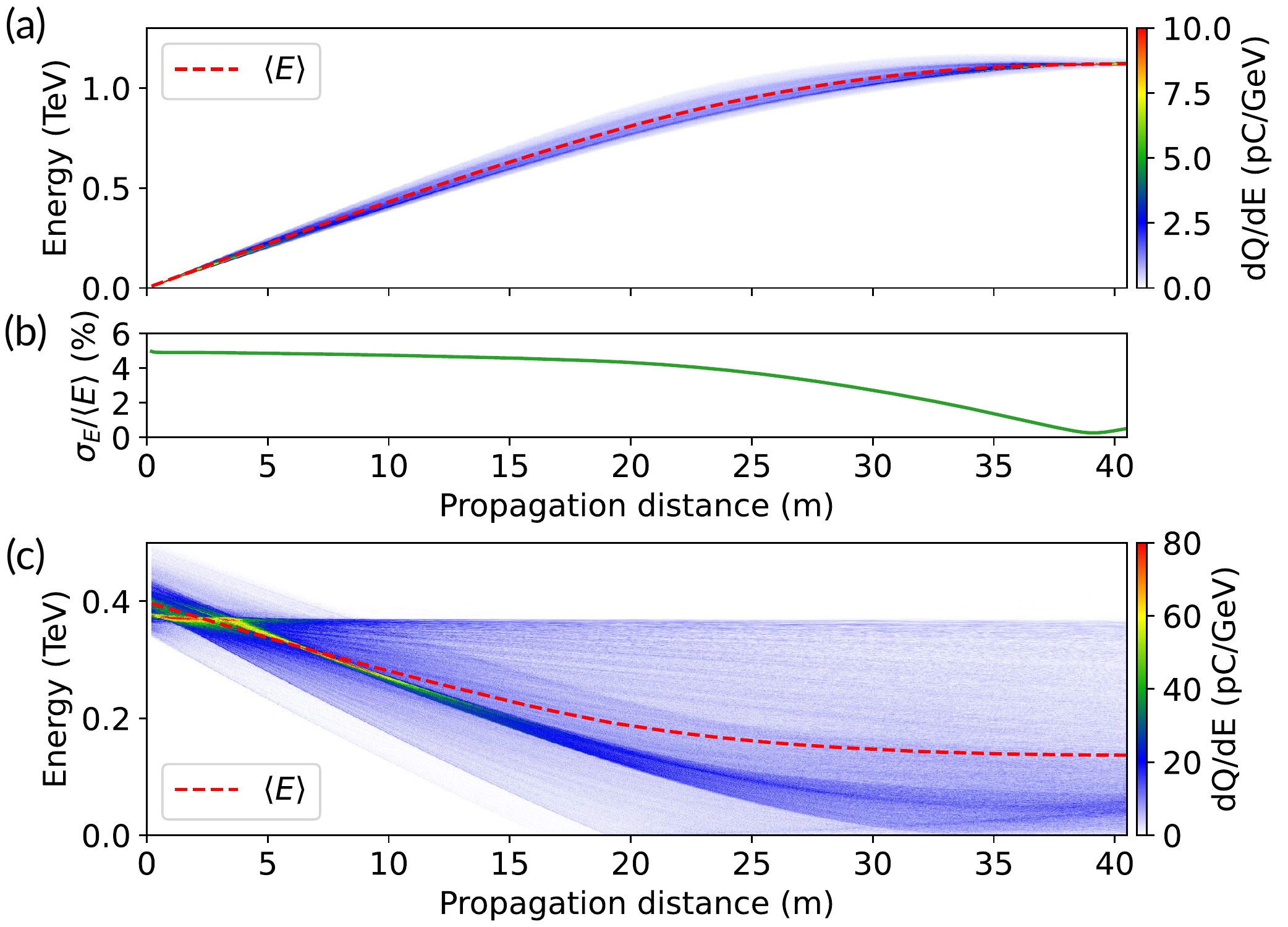}
\caption{Energy evolution along the second plasma stage. (a) Trailing bunch energy spectrum versus propagation distance, with charge-weighted mean shown as the dashed red line. The mean energy first grows linearly, then at a reduced rate, and eventually saturates near the plasma exit, reflecting the transition from non-evolving, underloaded to evolving, overloaded wake. (b) Relative rms energy spread of the 90\% core charge, dropping from $\sim$5\% to $\sim$0.3\% over the same distance. (c) Same as (a) for the drive bunch. A small fraction of the drive electrons reaches full depletion near 20 m and is subsequently re-accelerated to about 50 GeV in the last 20 m.}
\label{fig:spectra_evolution}
\end{figure}

The driver-to-trailing-bunch energy efficiency in this two-stage configuration is $\sim$3\%, well below the $\eta^2\approx16\%$ theoretical ceiling set by per-stage efficiencies of $\eta=0.4$ achievable for nonlinear PWFA~\cite{litos_high-efficiency_2014,joshi_plasma_2018}. The factor of $\sim$5 gap leaves room for optimization, for instance, increasing charge to 0.6 nC at the same final energy.

Figure \ref{fig:spectra_evolution} shows the evolution of the trailing bunch and driver energy spectra along the second stage (the acceleration part modeled using QPAD). Three phases are visible. In the first $\sim$20 m, the trailing bunch gains energy approximately linearly with an rms energy spread near 5\% [Figs. \ref{fig:spectra_evolution}(a),(b)] in an under-loaded, non-evolving wake. Near 20 m, a small fraction of the drive electrons reaches full depletion [Fig. \ref{fig:spectra_evolution}(c)], and the wake begins to change in shape and drop in amplitude. The partially depleted driver nevertheless continues to sustain a nonlinear wake, which now acts as an integrated energy dechirper~\cite{wu_phase_2019,shpakov_longitudinal_2019,darcy_tunable_2019}: the trailing bunch transitions dynamically from under- to over-loading the wake, rotating its longitudinal phase space and reducing the rms energy spread from 5\% at 20 m to $\sim$0.3\% at the plasma exit. Beyond 40 m, continued propagation would over-rotate the phase space and degrade both the mean energy and spread.

\begin{figure}[tb]
\centering
\includegraphics[width=\columnwidth]{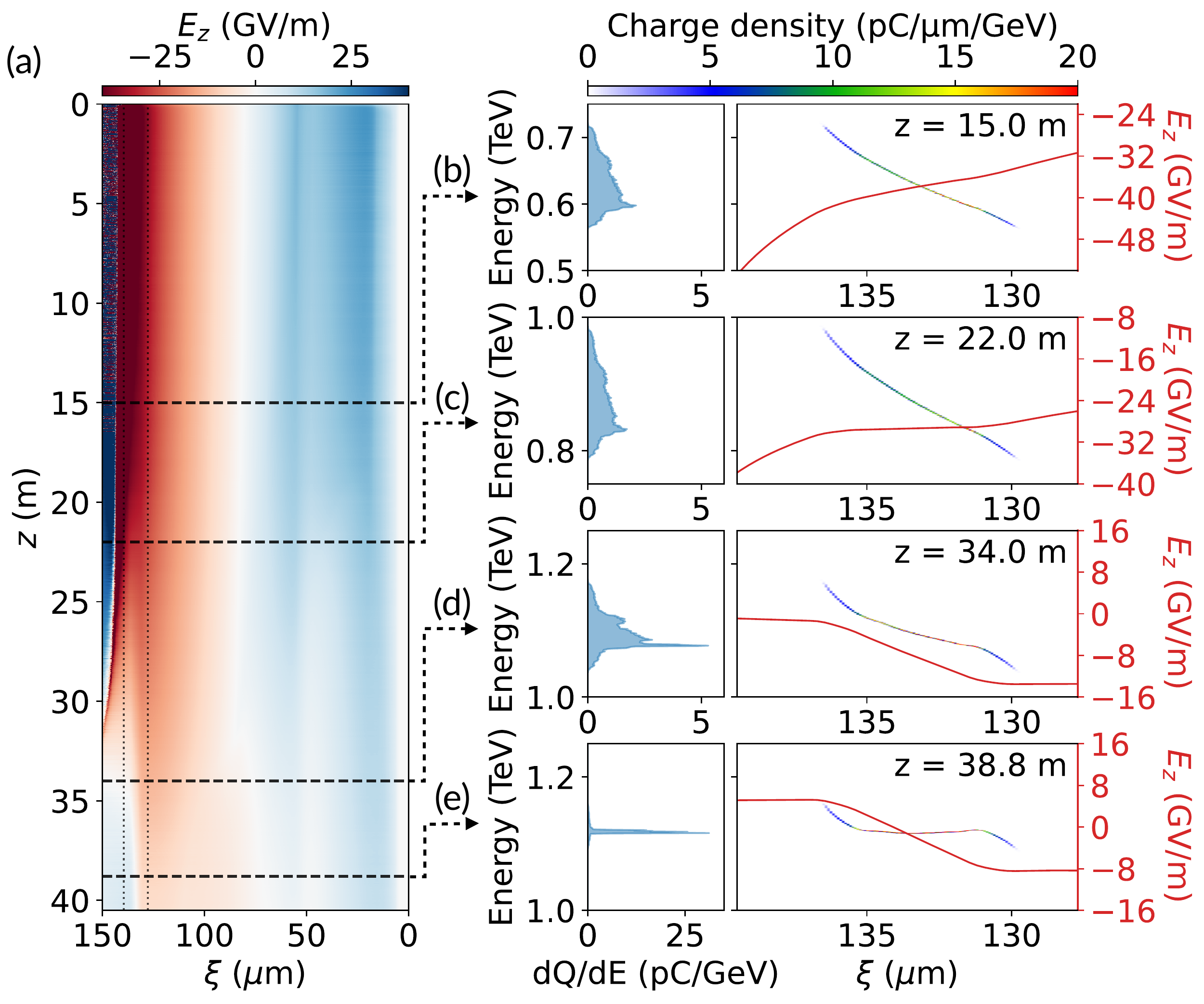}
\caption{Dynamic beam loading. (a) On-axis electric field of the wake as a function of propagation distance. The four dashed lines mark the locations at which the trailing bunch's longitudinal phase space (LPS) is evaluated, shown in (b)-(e). The trailing bunch first accumulates a large positive energy chirp in the underloaded wake (b), reaches optimal loading near $z = 22.0$~m (c), then overloads the wake and progressively rotates its LPS (d), until the LPS is eventually flattened at $z = 38.8$~m (e). By the exit, the rms energy spread is reduced to $\sim 0.3\%$.}
\label{fig:dynamic_beamloading}
\end{figure}

Figure \ref{fig:dynamic_beamloading} makes this dynamic beam loading (DBL) explicit by plotting the on-axis electric field [Fig.~\ref{fig:dynamic_beamloading}(a)] and the longitudinal phase space of the trailing bunch at four representative locations. The progression is continuous: in the first 20 m the trailing bunch underloads the wake (the $E_z$ field has a positive slope) and accumulates a large positive energy chirp [energy increasing from the front to the back, Fig.~\ref{fig:dynamic_beamloading}(b)]; near 22 m the wake (amplitude now reduced by partial pump depletion) is optimally loaded and the accelerating field flattens [Fig.~\ref{fig:dynamic_beamloading}(c)]; thereafter the trailing bunch overloads the wake ($E_z$ field now has a negative slope). The trailing bunch still gains energy [Fig.~\ref{fig:dynamic_beamloading}(d)] but the head of the beam gains energy more rapidly than the tail such that by 39 m the longitudinal phase space essentially flattens [Fig.~\ref{fig:dynamic_beamloading}(e)]. This progressive rotation reduces the projected rms energy spread from $\sim$5\% to $\sim$0.3\% [Fig. \ref{fig:spectra_evolution}(b)]. The plasma length is therefore a tunable design parameter for optimizing the DBL section to minimize the final energy spread; the energy spread remains below 0.5\% over a $\sim$3 m window around the optimum, providing several meters of tolerance for experimental adjustment.

This dynamic beam loading mechanism inverts the conventional PWFA design intuition. Most PWFA designs truncate the plasma before pump depletion to maintain a non-evolving wake; we instead extend past the depletion onset and let the evolving wake act as a built-in dechirper, which removes the trailing bunch energy chirp while increasing the energy gain by an additional $\sim$30\%, improving the efficiency and raising the energy transformer ratio to $\sim$3 despite a voltage transformer ratio of 2. Dynamic beam loading has been reported in LWFA~\cite{liu_scalable_2024} and in a single PWFA stage approaching pump depletion~\cite{dalichaouch_fully_2025}; here the wake is driven beyond the depletion onset---where a fraction of the drive electrons is fully depleted and re-accelerated---to reduce the energy spread of the final stage of a TeV-scale cascade.

Throughout the 40 m plasma, the trailing bunch's charge, peak current, and normalized emittance remain approximately constant at 0.12 nC, 8 kA, and $\sim$2.4 mm$\cdot$mrad, yielding a brightness of $\sim10^{15}~\rm A/m^2/rad^2$, consistent with previous downramp-injection measurements at similar plasma densities~\cite{zhang_plasma-wakefield_2025}. This brightness is four orders of magnitude higher than that of the stage 1 trailing bunch (2.4 kA, $\sim$160 mm$\cdot$mrad), confirming the brightness-reset property of PARC: the final beam brightness is set by self-injection in the last stage rather than by accumulated degradation across stages as in conventional concepts.

Hosing of the driver is the main instability concern in PARC, particularly in stage 1 where the long, shaped driver propagates 785 m in the high-transformer-ratio regime. Several betatron-detuning mechanisms have been identified that can mitigate hosing in PWFAs~\cite{mehrling_mitigation_2017-1,mehrling_suppression_2018,mehrling_mechanisms_2019,martinez_de_la_ossa_intrinsic_2018}; however, the high-transformer-ratio profile flattens $E_z$ along the drive bunch, which suppresses those mechanisms relying on energy-chirp or $E_z$-variation along the beam. Plasma ion motion provides an alternative mitigation channel that does not rely on the longitudinal wake structure: the ion motion induces a head-to-tail variation in the focusing force that damps the centroid oscillations on the timescale of a few betatron periods or less~\cite{mehrling_suppression_2018}. For the stage 1 driver in our simulation, the ion-motion parameter $\Lambda=Z_i(m/M_i)(I_b/I_A)(L_b/\sigma_x)^2\approx100$, where $\sigma_x\approx14~\mu$m is the ion-channel matched rms spot size~\cite{mehrling_suppression_2018}, falls in the relativistic ion-motion regime where hosing is suppressed within a single betatron period. The same parameter is $\Lambda\sim0.5$ for the stage 2 driver, for which ion motion damps the centroid oscillations within a few betatron periods. In addition, the stage 2 driver carries the large ($\sim$7\%) correlated energy spread imprinted for compression, which detunes the betatron frequency along the bunch and decoheres centroid oscillations within the first few meters of the 40 m stage~\cite{mehrling_mechanisms_2019}. Stable driver propagation also ensures stable trailing bunches, which are self-injected on the wake axis and inherit no centroid offset by construction---consistent with the absence of observable hosing in recent downramp-injection experiments~\cite{zhang_plasma-wakefield_2025}.

While beneficial for hosing suppression, ion motion can drive emittance growth via nonlinear focusing-field perturbations~\cite{an_ion_2017,benedetti_emittance_2017}. Nevertheless, for our stage 2 trailing bunch (the most demanding case owing to its small spot size and high charge density), the relevant ion-motion parameter $\Gamma=Z_i(m/M_i)(n_b/n_0)(k_pL_b)^2$~\cite{benedetti_emittance_2017} is $\sim0.2$ in hydrogen plasma, well below the $\Gamma\sim1$ threshold for noticeable growth, thus predicting a saturated emittance growth that is negligible.

A high-charge, intermediate-energy driver is the demanding element of this scheme. The simulation above uses a 45 GeV, 100 nC drive beam as an illustrative working point, but the architecture allows a broad parameter range; what matters is sufficient driver energy and charge to multiply through the cascade. Electron beam charges at the 100 nC level have been produced at lower energies in conventional linacs~\cite{conde_generation_1998}, and $\mu$C-class charges have been reported in laser-driven wakefield experiments~\cite{shaw_microcoulomb_2021}; extending such high-charge beams to the tens-of-GeV scale is the key driver development that PARC motivates. The brightness-reset property of PARC relaxes this requirement further: because the final beam brightness is set by self-injection in the last stage, the first-stage driver need not be high-current or low-emittance. A long, several-kA bunch driving a low-density plasma suffices, which simultaneously suppresses the beam-breakup (hosing) instability and CSR-induced emittance growth in the upstream accelerator.

Several building blocks of PARC have already been demonstrated experimentally. Early PWFA experiments at SLAC's Final Focus Test Beam (FFTB) used multi‑nC, picosecond‑scale electron bunches ($\sim$3 nC, $\sigma_z/c\approx2.3$ ps) to drive meter‑scale, low‑density ($\sim10^{14}~\rm cm^{-3}$) plasmas at peak accelerating gradients above 100 MV/m~\cite{muggli_meter-scale_2004}. Separately, an LWFA-PWFA hybrid scheme has shown, at the GeV scale, that an electron bunch from a laser wakefield accelerator can drive a subsequent PWFA stage that injects and accelerates a fresh trailing bunch with high efficiency~\cite{foerster_stable_2022,foerster_efficient_2026}. A full realization of PARC would therefore be the first demonstration of a hybrid platform that integrates conventional RF‑linac and plasma‑accelerator technologies. As a scaled, near‑term proof of concept, Appendix C shows that PARC applied to the EIC electron beam (10 GeV, 28 nC~\cite{willeke_electron_2021}) delivers 86 GeV bunches at 0.2 nC with $\sim0.8\%$ projected ($\sim0.02\%$ slice) energy spread.

In summary, we have introduced a new concept called Plasma Acceleration via Regenerative Cascading and shown through start-to-end PIC simulations that a 45 GeV, 100 nC driver produces a 1.1 TeV, 0.12 nC electron beam with $\sim$0.3\% energy spread and $\sim$2.4 mm$\cdot$mrad normalized emittance in two plasma stages with a total length less than one kilometer. By regenerating a fresh trailing bunch at each stage, PARC eliminates the alignment, synchronization, and matching tolerances that have limited conventional multi-stage schemes and enables multiplicative rather than additive energy gain. The sub-percent energy spread is obtained through dynamic beam loading past pump depletion, and the final beam brightness is reset by self-injection in the last stage. These results bring TeV-scale electron beams within reach of a single high-charge driver, opening a path to energy-frontier colliders and to SFQED in the $\chi\gg1$ regime.

\begin{acknowledgments}
\textit{Acknowledgments---}This work was supported by the U.S. Department of Energy through Grant No. DE-SC0010064. Simulations were performed using resources of the National Energy Research Scientific Computing Center (NERSC), a U.S. DOE Office of Science User Facility located at Lawrence Berkeley National Laboratory, operated under Contract No. DE-AC0205CH11231 using NERSC Award HEP-ERCAP-MP113. We thank Professor Warren Mori’s group for the PIC codes used in this work.
\end{acknowledgments}

\textit{Data availability---}The data are available from the authors upon reasonable request.

%============================================================================
% REFERENCES
%============================================================================
\bibliographystyle{apsrev4-2}
\bibliography{PARC}

%============================================================================
% APPENDICES
%============================================================================
\onecolumngrid
\section*{End Matter}
\twocolumngrid

\textit{Appendix A: Simulation details---} All stages are modeled in two sections: downramp injection and pre-acceleration to above 100 MeV using OSIRIS, followed by long-distance acceleration using QPAD. Both codes use a quasi-3D (cylindrical) geometry retaining only the $m=0$ azimuthal mode, with 8 macroparticles per cell along the azimuthal direction. The $m=0$ mode is sufficient here because the beams are azimuthally symmetric and the self-injected trailing bunch is on-axis by construction; hosing, an intrinsically $m\geq1$ phenomenon, is discussed in the main text. Each plasma profile consists of a 50 $c/\omega_p$ entrance ramp, a 50 $c/\omega_p$ plateau, a 10 $c/\omega_p$ downramp for injection, a uniform region of variable length at density $n_0$, and a 1200 $c/\omega_p$ exit ramp. All ramps follow a $\sin^2$ shape that connects different densities smoothly. The plasma is represented by $4\times4\times8$ ($z\times r\times\theta$) particles per cell with a 2 eV finite electron temperature, which suppresses artificially high current spikes that appear in the injected bunch for zero-temperature plasma. Across all simulations only $n_0$ and the downramp ratio vary. Unless otherwise noted, projected energy spreads, emittances, and charges of accelerated trailing bunches refer to the 90\% core charge of the bunch; current profiles use the full distribution after removing halo particles, and slice quantities are evaluated for the 90\% core of the charge in each slice.

In stage 1 of the main result, OSIRIS uses a co-moving window of $18 \times 6\;c/\omega_p$ ($z \times r$), $\Delta z = \Delta r = (1/64)\;c/\omega_p$, and time step $dt = (1/128)\;\omega_p^{-1}$, with $\omega_p$ evaluated at $n_0 = 4\times10^{13}~\text{cm}^{-3}$. The shaped driver ($\sigma_r = 92\;\mu\text{m}$, $\varepsilon_n = 50\;\text{mm}\cdot\text{mrad}$, $\beta = 15\;\text{m}$) is represented by $2\times2\times8$ particles per cell. The trailing bunch is injected and accelerated for 420 $c/\omega_p$ to $\sim$180 MeV. The 6D phase space of both the trailing bunch and driver is then exported to QPAD, which retains the same window and resolution with $dt=40~\omega_p^{-1}$ and $8\times8\times8$ particles per cell.

The compressed stage 1 trailing bunch is imported back into OSIRIS for stage 2 and self-focuses in the entrance ramp without external focusing optics. Stage 2 uses peak density $n_0=10^{17}~\text{cm}^{-3}$ and a $12\times6~c/\omega_p$ moving window; resolution and plasma parameters are otherwise identical. A new trailing bunch is injected and accelerated by 420 $c/\omega_p$ to $\sim$0.3 GeV, and both beams are then exported to QPAD for 40 m of acceleration. The self-injected trailing bunch is born nearly matched: residual mismatch and its correlated energy spread increase the projected emittance by less than a factor of two within the first $\sim$0.1 m, after which it remains
constant. Because the driver partially depletes in stage 2, low-energy depleted electrons can have under-resolved betatron motion at large time steps. We therefore tested convergence by varying the time step: the final trailing bunch energy is 1.18 TeV (0.2\% spread) at $dt=40~\omega_p^{-1}$ versus 1.12 TeV (0.3\%) at $dt=20~\omega_p^{-1}$. The main text reports the $dt=20~\omega_p^{-1}$ case.

For the EIC case study (Appendix C), the EIC beam is compressed to 9.5 ps full length and shaped to the same current profile as in Fig. \ref{fig:stages}(a), focused to $\sigma_r=118~\mu\text{m}$ with $\varepsilon_n=400$ mm$\cdot$mrad and $\beta=0.69$ m. Stage 1 uses a 3:1 ratio density downramp with $n_0=5\times10^{14}~\text{cm}^{-3}$ over $\sim$55 m total plasma length; stage 2 uses a 2:1 ratio density downramp at $n_0=10^{17}~\text{cm}^{-3}$ over 4 m. Window size, resolution, particles per cell, and time steps are identical to the main-result simulations.

\begin{figure}[tb]
\centering
\includegraphics[width=\columnwidth]{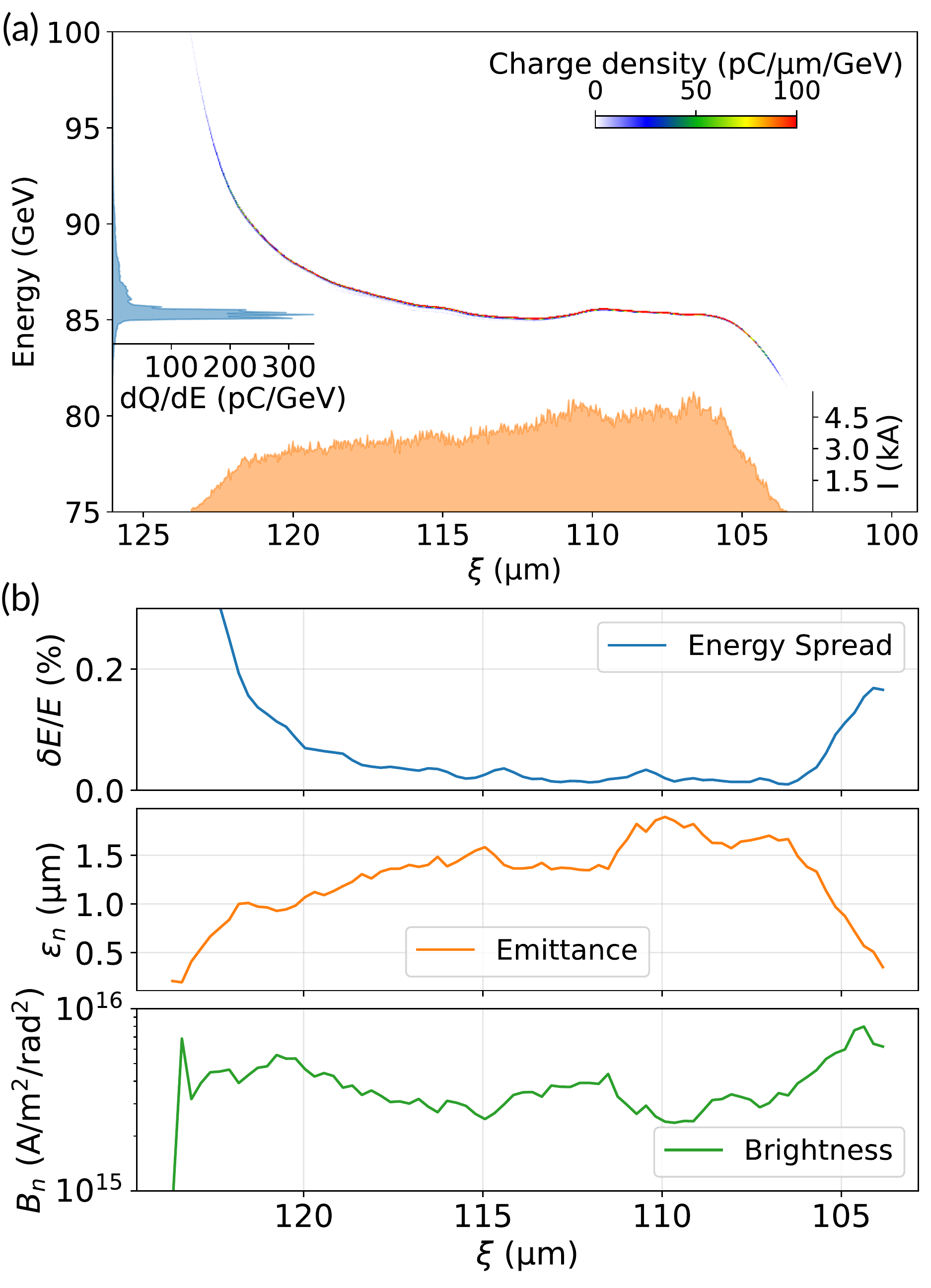}
\caption{PARC applied to the EIC electron beam (10 GeV, 28 nC); simulation parameters are given in Appendix A. (a) Longitudinal phase space of the final trailing bunch, with projected energy spectrum (left) and current profile (bottom): 86 GeV mean energy, 0.2 nC charge, $\sim$0.8\% projected rms energy spread, $\sim$5 kA peak current. (b) Slice-resolved beam parameters: relative energy spread (top), normalized emittance (middle), and brightness (bottom, log scale). Across the core, the bunch has a $\sim$0.02\% slice energy spread, $\sim$1.4 mm$\cdot$mrad slice emittance, and $\sim4\times10^{15}~\rm{A/m^2/rad^2}$ brightness.}
\label{fig:EIC}
\end{figure}

\textit{Appendix B: Inter-stage trailing bunch compression---} At the exit of stage 1, the spent driver is discarded and the trailing bunch is compressed analytically. The stage 1 trailing bunch exits with 400 GeV mean energy, 2.3 nC total charge, $\sigma_z\approx120\;\mu$m, $\varepsilon_n \approx 160\;\text{mm}\cdot\text{mrad}$, and a nearly linear chirp $h = 5.91\times10^{-4}\;\mu\text{m}^{-1}$ ($\sim$7\% correlated spread). Compression via a four-dipole chicane with $z_f = z_i + R_{56}\delta + T_{566}\delta^2$ with $R_{56} = -0.91/h \approx -1.54$ mm yields $C \approx 11$ and $\sigma_{z,f} \approx 14\;\mu$m. At 400 GeV, a 0.8~T chicane with 14~m dipoles and 2~m drifts provides the required $R_{56}$ in $\sim$63~m total length. The high beam energy and small $|R_{56}|$ keep both the CSR-induced energy spread (MeV-level) and the emittance growth negligible, so the final bunch length is set by the chirp alone. This treatment was further verified by tracking the stage 1 trailing bunch through a realistic quadrupole-and-chicane beamline with \textsc{Ocelot}~\cite{agapov_ocelot_2014}: the tracked and analytic compression agree at the percent level, and a stage 2 simulation restarted from the tracked phase space reproduces similar results (1.04 TeV, 0.38\% energy spread).

\textit{Appendix C: Case study using the EIC beam---} The Electron-Ion Collider (EIC) under construction is designed to provide 10 GeV, 28 nC electron beams at up to 1 Hz repetition rate. This beam could be extracted either via a dedicated beamline before storage-ring injection or out from the storage-ring once polarization drops below 70\%. Either path could serve as a near-term, facility-scale driver for PARC. Figure \ref{fig:EIC} shows a PIC simulation in which the compressed and shaped EIC beam drives two PARC stages (parameters are given in Appendix A). The final trailing bunch reaches 86 GeV mean energy, 0.2 nC charge, $\sim$0.8\% projected (0.02\% slice) rms energy spread, $\sim$1.4 mm$\cdot$mrad normalized emittance, and $\sim$5 kA peak current. The slice brightness ($\sim4\times10^{15}~\rm A/m^2/rad^2$) is comparable to that of the 1.1 TeV main result, demonstrating brightness reset across more than an order of magnitude in driver energy. Such a beam would enable fixed-target nuclear physics measurements beyond current capabilities and, when collided with an intense laser pulse, access strong-field QED regimes inaccessible to existing facilities.

\end{document}